\documentclass[11pt,a4paper,nofootinbib,superscriptaddress,aps]{revtex4-2}
\usepackage{graphicx,float,wrapfig,subfig}
\usepackage{amsfonts,amsmath,amssymb,amstext}
\usepackage{latexsym}
\usepackage{bm}
\usepackage{color}
\usepackage{xcolor}
\usepackage[normalem]{ulem}
\usepackage{MnSymbol}
\usepackage[colorlinks=true,linktocpage=true,linkcolor=blue,citecolor=blue]{hyperref}
\usepackage{slashed}
\usepackage{verbatim}
\usepackage{braket}
\usepackage{orcidlink}   
\usepackage{cancel}

\def\be {\begin{equation}}
\def\ee {\end{equation}}
\def\bea {\begin{eqnarray}}
\def\eea {\end{eqnarray}}

\begin{document}

\title{Electromagnetic response of a relativistic drifting plasma}	

	\author{Ashes Modak}
	\email{22dr0061@iitism.ac.in}
	\affiliation{Department of Physics, Indian Institute of Technology (Indian School of Mines) Dhanbad, Jharkhand 826004, India.}

    \author{Anowar Shaikh}
	\email{20004634@iitb.ac.in}
	\affiliation{Department of Physics, Indian Institute of Technology Bombay, Mumbai 400076, India.}

	\author{Manu Kurian}
	\email{manukurian@iitism.ac.in}
	\affiliation{Department of Physics, Indian Institute of Technology (Indian School of Mines) Dhanbad, Jharkhand 826004, India.}
  
	\author{Binata Panda}
	\email{binata@iitism.ac.in}
	\affiliation{Department of Physics, Indian Institute of Technology (Indian School of Mines) Dhanbad, Jharkhand 826004, India.}

    \author{Sadhana Dash}
	\email{sadhana@phy.iitb.ac.in}
	\affiliation{Department of Physics, Indian Institute of Technology Bombay, Mumbai 400076, India.}
	
\begin{abstract}
We investigate the charge transport properties of a relativistic drifting plasma using the kinetic theory within the relaxation time approximation.  The collective drift induced by electromagnetic fields is described in terms of a suitably modified distribution function. The analysis is done for both constant and time-dependent field configurations. For constant electromagnetic fields, we obtain the Hall drift current that arises from the transverse motion of charged particles in electric and magnetic fields. Extending the framework to time-dependent electric fields, we show that their temporal variations give rise to polarization drift, which significantly alters the structure of the induced current and introduces additional components along both the conventional drift and polarization directions. We present a quantitative estimate of the Hall drift and polarization-induced contributions in the quark–gluon plasma and study the temperature dependence of the associated charge transport coefficients in the QCD.
\end{abstract}
\maketitle 
	\section{Introduction}
Understanding the behavior of relativistic plasmas under the influence of electromagnetic fields is crucial for many physical systems, ranging from laboratory experiments to astrophysical environments. A particularly important example is the electromagnetic response of the hot and dense quark–gluon plasma (QGP) produced in relativistic heavy-ion collision experiments \cite{Grasso:2000wj,Giovannini:2003yn, Kharzeev:2015kna, Charbonneau:2009ax, Ohnishi:2014uea}. Extremely strong electromagnetic fields are expected to be generated during the early stages of these collisions~\cite{Tuchin:2013apa, Yan:2021zjc}. These fields can significantly influence the thermodynamic properties and transport behavior of the medium, thereby affecting its subsequent evolution. Various approaches have been employed to compute the transport coefficients of the QGP in the presence of electromagnetic fields \cite{Rath:2019vvi, Rath:2020idp, Rath:2018tdu, Kurian:2018qwb, Hattori:2016lqx, Kurian:2019fty, Puglisi:2014sha,  Ghosh:2024owm, Shaikh:2024tjn, Jiang:2024jxq, Astrakhantsev:2019zkr,Dwibedi:2025xho}. The induced current densities due to the electromagnetic fields in the medium provide a direct link between the microscopic dynamics of charge carriers and the macroscopic evolution of electromagnetic fields. The study of electromagnetic responses of a chiral plasma has also gained attention over a last few years. The chiral asymmetry of the medium coupled with the electromagnetic fields generated during heavy-ion collisions can lead to anomalous transport effects, such as the Chiral Magnetic Effect (CME), Chiral Separation Effect (CSE), etc. \cite{Kharzeev:2013ffa, Kharzeev:2007jp, Fukushima:2008xe, Kharzeev:2007tn, Metlitski:2005pr, Jensen:2013vta, Ghosh:2023ghi}.

Transport properties of these systems are highly sensitive to both initial conditions and the temporal changes in electromagnetic fields \cite{Tuchin:2013ie, Bzdak:2011yy, Singh:2023pwf}. Analyzing the charge transport in a plasma in the presence of a general time-varying electromagnetic field is a complex task. Electromagnetic responses of a relativistic plasma can be studied in different approaches, depending upon the collision regime. In the case where neutral particles interact frequently with charged particles (when the collision rate is high), both electric and magnetic fields can be treated as small perturbations~\cite{Chen:2016xtg,Feng:2017tsh,Gowthama:2020ghl,K:2021sct,Rath:2025pho,K:2022pzc,Kurian:2020qjr, Dey:2020sbm,Rath:2025zhw}. Although the charged particle dynamics will be directly influenced by the electromagnetic fields,  frequent collisions with neutral particles exert a strong drag on the system such that one can use the Fermi-Dirac (FD) distribution function as the equilibrium distribution. For the regime where interaction of neutral particles with charged particles is relatively weak, the standard FD distribution may not be the best choice for the zeroth order distribution function. This is because the charge particles in the plasma experiences a drift and can capture this effect by a modified distribution function \cite{Gorbar:2016qfh}. A particularly important situation arises when electric and magnetic fields are mutually perpendicular. In this case, charged particles execute rapid gyro-motion around a guiding center, while the guiding center itself acquires a drift velocity orthogonal to both fields. In a relativistic plasma, this drift is described by the standard $\mathbf{E}\times\mathbf{B}$ velocity as \cite{chen2016introduction, krall1973principles},
\begin{align}\label{normal_drift}
\mathbf{v}_d \;=\; \frac{\mathbf{E}\times\mathbf{B}}{B^2}\,,
\end{align}
with \(c=1\) and \(E_\perp<B\), ensuring \(|\mathbf{v}_d|<1\). An instructive way to view this state is through a change of reference frame: in the co-moving frame that drifts with \(\mathbf{v}_d\), the transverse electric field disappears under a Lorentz transformation, and the plasma can be described by an equilibrium FD distribution. Transforming back to the laboratory frame then yields a boosted equilibrium distribution whose exponent contains the characteristic shift \(\epsilon-\mathbf{p}\cdot\mathbf{v}_d\), providing a convenient starting point for the study of transport process in a drifting medium.

In the present study, we analyze the electric charge transport in a relativistic drifting plasma. The current densities are derived by solving the relativistic Boltzmann equation within the relaxation time approximation (RTA). We linearize around the drift-modified equilibrium distribution and systematically compute the non-equilibrium correction to the distribution function. Another ubiquitous feature of evolving plasmas is inhomogeneity. Even at the near-equilibrium state, such inhomogeneity can be parametrized through gradients of space-time-dependent thermodynamic fields. In a drifting medium, these inhomogeneities couple to the boosted distribution and generate additional, physically meaningful contributions to both the charge density and current density. For quantitative estimates, we present numerical results using parameter values appropriate for the QGP medium. The analysis has been carried out for two field configurations. In \emph{Case I}, we consider constant electric and magnetic fields, recovering the baseline drifting response of the medium. In \emph{Case II}, we consider a time-dependent electric field along with the magnetic field, which generates both a conventional Hall drift current  and a distinct current component associated with polarization drift that arises due to the temporal variation of the electric field. 

The paper is organized as follows. In Sec.~\ref{RBTE}, we present the formalism of electric charge transport in a relativistic drifting plasma and derive the associated current densities for different electromagnetic field configurations. In Sec. \ref{section III}, we quantitatively analyze the results for the QGP medium and we summarize the study in Sec.~\ref{summary}. Detailed description on the expansion of drift-modified distribution function and mathematical forms of momentum integrals in both massive and massless limit are discussed in the appendices.

\noindent \textbf{Notations and conventions:} Throughout this work, $\partial_\mu$ denotes differentiation with respect to the space-time coordinate $x^\mu$, whereas $\partial^{(p)}_\mu$ denotes differentiation with respect to the four-momentum $p^\mu$. In the local rest frame, the fluid four-velocity is chosen as $u^\mu=(1,0,0,0)$ and satisfies the normalization condition $u^\mu u_\mu = 1$. The index $i$ labels quark flavors; for the quark-gluon plasma (QGP), it corresponds to the up ($u$), down ($d$), and strange ($s$) quarks. The quantities $q_i$, $g_i$, and $\delta f_i$ ($\delta \bar f_i$) denote, respectively, the electric charge, degeneracy factor, and infinitesimal deviation of the quark (antiquark) distribution function for the $i^{\text{th}}$ flavor. The symbol $m_i$ represents the current quark mass, taken as 3 MeV, 5 MeV, and 100 MeV for the up, down, and strange quarks, respectively. The gluonic degeneracy is given by $g_g = 2(N_c^2-1)$, while the quark and antiquark degeneracies for each flavor are $g_f = g_{\bar f} = 2N_c$, where the number of colors is fixed as $N_c=3$.
\section{Electric charge transport in a drifting plasma}\label{RBTE}
The electric four-current of a drifting plasma can be defined in terms of particle momentum distribution
function as follows,
\begin{align}\label{current}
   \mathbf{J}^\mu 
           = \sum_{i} g_{i} q_i \int \frac{d^{3}p}{(2\pi)^{3}\epsilon_{i}}{\mathbf{p}^\mu}f^{(v_d)}_{\mathrm{eq},i}(t,x,p),
\end{align}
with $\epsilon$ as the energy density. In the presence of external electromagnetic fields, the single-particle distribution function deviates slightly from local equilibrium and can be written as, 
\begin{align}\label{11}
f^{(v_d)}(t,x,p) = f_{\mathrm{eq}}^{(v_d)}(t,x,p) + \delta f(t,x,p),
\end{align}
where \( f_{\mathrm{eq}}^{(v_d)} \) denotes the equilibrium distribution function of the charged particles in a drifting plasma \footnote{In general, under local equilibrium both the temperature and the chemical potential may depend on the space-time coordinates. In the present analysis, however, we neglect the space-time dependence of the temperature and retain only the possible space-time dependence of the chemical potential \cite{Gorbar:2016qfh}.}and \( \delta f \) represents a small perturbation satisfying \( |\delta f| \ll f_{\mathrm{eq}}^{(v_d)} \). 
The first step in the estimation of the current density is to obtain the equilibrium and non-equilibrium part of the distribution. Charged particles in the presence of electromagnetic fields experience drift, which in turn, can modify their momentum distribution function. To that end, the choice of field configurations plays an important role. In this study, we consider two different cases. 
\subsection*{Case I: $\mathbf{E}\times \mathbf{B}$ drift}
In general, the momentum distribution of fermionic particles at local thermodynamic equilibrium can be described by the standard FD  distribution function  \cite{Kremer2014TheoryAA}, 
\begin{align}
f_{\mathrm{eq}}^{\mathrm{std}} = \frac{1}{\exp\!\left[{\beta(p^{\mu}u_{\mu} - \mu)}\right] + 1},
\end{align}
where $u^{\mu}$ denotes the local four-velocity of the medium, $\mu$ is the local chemical potential, and $\beta = 1/T$, with $T$ representing the temperature. In the presence of electromagnetic fields, charged particle experience $\mathbf{E}\times\mathbf{B}$ drift depending upon the relative orientation of the electric and magnetic fields, as described in Eq.~(\ref{normal_drift}). In this scenario, it is useful to distinguish between two reference frames: the laboratory frame $(F)$ and the drifting (or co-moving) frame $(F')$. The laboratory frame $(F)$ is the frame in which the external electromagnetic fields $\mathbf{E}$ and $\mathbf{B}$ are applied and measured. In this frame, charged particles experience both fields simultaneously and respond by acquiring a collective drift velocity $\mathbf{v}_d$. This drift represents the bulk motion of the plasma due to the combined action of the electric and magnetic fields. In contrast, the drifting or co-moving frame $(F')$ moves with the same velocity $\mathbf{v}_d$ relative to the laboratory frame. When viewed from this frame, the electric field effectively disappears under a Lorentz transformation, leaving only a magnetic field \cite{LandauLifshitzCTF}. 
Consequently, the plasma in this co-moving frame is in perfect local thermodynamic equilibrium and described by distribution function as,
\begin{align}
f'_{\mathrm{eq}} = \frac{1}{\exp[\beta'(p^{'\nu}u^{'}_{\nu}-\mu^{'})]+1}=\frac{1}{\exp[\beta'(\epsilon^{'}-\mu^{'})]+1},
\end{align}
where $\epsilon'=\sqrt{\mathbf{p}'^{\,2}+m^{2}}$ denotes the single-particle energy, $\mu'$ is the chemical potential, $\beta'=1/T'$ is the inverse temperature.
In this frame, the four-momentum and the fluid four-velocity are given by $p'^{\mu} = (\epsilon',\, \mathbf{p}')$ and $u'^{\nu} = (1,\,0,\,0,\,0)$, respectively. To express the equilibrium distribution function in the laboratory frame, we perform a Lorentz transformation from the co-moving (drifting) frame. 
The four-velocity of the plasma in the laboratory frame is given by 
$u^{\mu} = (u^{0}, \mathbf{u}) = \left({1}/{\sqrt{1 - ({v}_d)^{2}}},{\mathbf{v}_d}/{\sqrt{1 - ({v}_d)^{2}}}\right)$, 
while the temperature and chemical potential transform as 
$T = T'\sqrt{1 - ({v}_d)^{2}}$ 
and 
$\mu = \mu'\sqrt{1 - ({v}_d)^{2}}$, 
respectively. Hence, the drift-modified equilibrium distribution function for particle and antiparticle can be written as follows~\cite{Gorbar:2016qfh}
\begin{align}\label{Particle Distribution Function}
&f_\mathrm{eq}^{(v_d)} = 
\frac{1}{\exp\left[ \beta \left( \epsilon - \mathbf{p} \cdot \mathbf{v}_d - \mu \right) \right] + 1}, \\
& \bar{f}_\mathrm{eq}^{(v_d)} = 
\frac{1}{\exp\left[ \beta \left( \epsilon - \mathbf{p} \cdot \mathbf{v}_d + \mu \right) \right] + 1},\label{AntiParticle Distribution Function}
\end{align}
with $\epsilon = \sqrt{\mathbf{p}^{2} + m^{2}}$ as the single-particle energy, $\mu$ as the chemical potential, and $\beta = 1/T$ denotes the inverse temperature in the laboratory frame. The term $\mathbf{p}\cdot\mathbf{v}_d$ arises from the Lorentz transformation of the particle energy between the co-moving and laboratory frames, leading to a boosted distribution function that describes the equilibrium state of a  drifting plasma. 

The next step is the estimation of the non-equilibrium part of the distribution function. The evolution of $f(t,x,p)$ is described by the relativistic Boltzmann transport,
\begin{align}
p^{\mu} \, \partial_{\mu} f(t,x,p)
+ q \, F^{\mu\nu} p_{\nu} \,{\partial_{\mu}^{(p)} f(t,x,p)}
= C[f],
\end{align}
where $F^{\mu\nu}$ is the electromagnetic field-strength tensor and $C[f]$ denotes the collision integral that encodes the microscopic interactions responsible for driving the system toward equilibrium. We employed the widely used RTA framework to incorporate the collisional effect. In this framework, the collision term is substituted
with a relaxation term of the following form \cite{ANDERSON1974466},
\begin{align}
C_{\mathrm{RTA}}[f] = -\frac{1}{\tau}\,(f^{(v_d)} - f^{(v_d)}_\mathrm{eq}),
\end{align}
with $\tau$ as the thermal relaxation time. It is important to distinguish between the roles of different components of the electric field in the present analysis. The drift motion of charged particles arises from the transverse component of the electric field, $i.e.$, $\mathbf{E}_\perp \perp \mathbf{B}$, which leads to the well-known $\mathbf{E}\times\mathbf{B}$ drift.
In contrast, the deviation from local equilibrium is driven by the longitudinal component of the electric field, $\mathbf{E}_\parallel \parallel \mathbf{B}$. Therefore, while $\mathbf{E}_\perp$ determines the drift kinematics, $\mathbf{E}_\parallel$ is responsible for driving the system out of equilibrium and governs the transport response. By employing order-by-order expansion method, we can further simplify the Boltzmann equation as, 
\begin{align}\label{Particle RBTA}
\frac{\partial f^{(v_d)}}{\partial t}
+\Bigg\{q\,\mathbf{E}+q\,(\mathbf{V}\times\mathbf{B})\Bigg\}\cdot
\frac{\partial f^{(v_d)}}{\partial \mathbf{p}}
+\mathbf{V}\cdot
\frac{\partial f^{(v_d)}}{\partial \mathbf{x}}
=
-\frac{1}{\tau}\left(f^{(v_d)}-f_{\mathrm{eq}}^{(v_d)}\right).
\end{align}
Here, $V=\frac{p}{\epsilon}$ is the particle velocity. Using Eq.~\eqref{11}, we solve Eq.~\eqref{Particle RBTA} to obtain the first-order correction to the distribution function for charged particles drifting with velocity $\mathbf{v}_d$, given by
\begin{align}\label{deltaf}
\delta f^{(1)}
=-
\tau\,\frac{D_{\mathrm{eq}}^{(v_d)}}{T}
\left\{
\frac{\partial(\mu+\mathbf{p}\cdot\mathbf{v}_d)}{\partial t}
-
\frac{q\,(\mathbf{E}\cdot\mathbf{B})(\mathbf{V}\cdot\mathbf{B})}{B^{2}}
+
\mathbf{V}\cdot
\frac{\partial(\mu+\mathbf{p}\cdot\mathbf{v}_d)}{\partial \mathbf{x}}
\right\},
\end{align}
where $D_{\mathrm{eq}}^{(v_d)}
=
f_{\mathrm{eq}}^{(v_d)}\left(1-f_{\mathrm{eq}}^{(v_d)}\right)$.
Similarly, one can obtain the non-equilibrium part of the antiparticle as \footnote{For antiparticles, the electric charge and the chemical potential are taken with opposite signs.},
{\begin{align}\label{deltafabar}
\delta \bar f^{(1)}
=- \tau\,\frac{\bar D^{(v_d)}_{\rm eq}}{T}
\left\{
\frac{\partial\!\left(\mathbf{p}\!\cdot\!\mathbf{v}_d-\mu\right)}{\partial t}
-\frac{\bar q\,(\mathbf{E}\!\cdot\!\mathbf{B})\,(\mathbf{V}\!\cdot\!\mathbf{B})}{B^{2}}
+\mathbf{V}\cdot\frac{\partial\!\left(\mathbf{p}\!\cdot\!\mathbf{v}_d-\mu\right)}{\partial \mathbf{x}}
\right\},
\end{align}
where $\bar D^{(v_d)}_{\rm eq}=\bar f^{(v_d)}_{\rm eq}\left(1-\bar f^{(v_d)}_{\rm eq}\right)$.

The current density of the drifting plasma can be obtained by substituting Eq.~(\ref{Particle Distribution Function}) and Eq.~(\ref{deltaf}) in Eq.~(\ref{current}). We obtain, 
\begin{align}\label{J Total}
\mathbf{J}^{(\text{Total})} &= \sum_{i}g_i q_i
\Bigg\{\int \frac{d^{3}p}{(2\pi)^{3}} 
\mathbf{V} f_{\mathrm{eq},i}^{(v_{d})}
+ \int \frac{d^{3}p}{(2\pi)^{3}}\frac{\tau D_{\mathrm{eq},i}^{(v_d)}}{T}
\mathbf{V} q \frac{(\mathbf{E} \cdot \mathbf{B})(\mathbf{V} \cdot \mathbf{B})}{B^2}- \int \frac{d^{3}p}{(2\pi)^{3}}\mathbf{V} \left( \mathbf{V} \cdot \frac{\partial \mu}{\partial \mathbf{X}} \right)\frac{\tau D_{\mathrm{eq},i}^{(v_d)}}{T}\nonumber\\
&-\int \frac{d^{3}p}{(2\pi)^{3}} \mathbf{V} \left( \mathbf{V} \cdot \frac{\partial (\mathbf{p} \cdot {\mathbf{v}_d})}{\partial \mathbf{X}} \right)\frac{\tau D_{\mathrm{eq},i}^{(v_d)}}{T}-\int \frac{d^{3}p}{(2\pi)^{3}}\mathbf{V} \frac{\partial \mu}{\partial t}\frac{\tau D_{\mathrm{eq},i}^{(v_d)}}{T}
- \int \frac{d^{3}p}{(2\pi)^{3}}\mathbf{V} \frac{\partial (\mathbf{p} \cdot \mathbf{v}_{d})}{\partial t}\frac{\tau D_{\mathrm{eq},i}^{(v_d)}}{T}
\Bigg\}\nonumber\\
&=\mathbf{J}^{(Hall)}+\mathbf{J}^{(1)}+\mathbf{J}^{(2)}.
\end{align}
The resulting current in a drifting plasma consists of both equilibrium and non-equilibrium components. The equilibrium contribution, referred to here as the Hall current, arises from the steady drift of charged particles in the electromagnetic field background, whereas the non-equilibrium terms originate from temporal and spatial variations of the chemical potential, drift velocity, and relaxation dynamics. To systematically analyze the individual contributions, the distribution function is expanded in a Taylor series around its equilibrium form in powers of the drift velocity $\mathbf{v}_d$. Carrying out the momentum integration, we obtain the Hall current as \footnote{The general structure of the momentum integrals is discussed in detail in Appendix~\ref{General Momentum Integrals and Bessel Function Structure}.}
\begin{align}\label{J Hall}
    \mathbf{J}^{(Hall)}=\sum_{i} g_iq_i\Bigg\{\mathbf{v}_{d}\frac{m_i^{2}T}{\pi^{2}}\sum_{j=1}^{\infty}\frac{(-1)^{j+1}}{j}\sinh{\left(\frac{j\mu}{T}\right)}K_{2}\left(\frac{jm_i}{T}\right)\Bigg\}.
\end{align}
It is worth noting that the resulting current is proportional to the ${\bf E}\times{\bf B}$ drift in the medium.  Similarly, we obtain $\mathbf{J}^{(1)}$ as follows
\begin{align}\label{J1}
    \mathbf{J}^{(1)}=\sum_{i} g_iq_i^2\Bigg\{\frac{\tau}{T}\frac{(\mathbf{E}\cdot\mathbf{B})(\mathbf{B}\cdot \hat{\mathbf{v}}_d)}{B^{2}}\left(H_{1,i}+v_d^2H_{2,i}\right)\Bigg\}\hat{\mathbf{v}}_d.
\end{align}
This contribution $\mathbf{J}^{(1)}$ vanishes when the electric and magnetic fields are mutually perpendicular. Its overall structure ensures that the induced current remains aligned with the drift-velocity direction, reflecting the interplay between the electric and magnetic fields in determining the transport response. In addition to this, there is a component associated with the inhomogeneity of the plasma, $\mathbf{J}^{(2)}$. We obtain,
\begin{align}\label{J2}
    \mathbf{J}^{(2)}=&-\sum_{i} g_iq_i\Bigg\{\Big(\hat{\mathbf{v}}_d\cdot \frac{\partial \mu}{\partial \mathbf{X}}\Big)\frac{\tau}{T}\left(H_{1,i}+v_d^2H_{2,i}\right) \hat{\mathbf{v}}_d+\frac{1}{5\pi^{2}}\frac{\tau}{T}\mathbf{v}_d(\mathbf{\nabla}\cdot \mathbf{v}_d)H_{3,i}   
    +\mathbf{v}_d\,\,\tau\frac{\partial\mu}{\partial t}H_{4,i}\nonumber\\
    &+\hat{\mathbf{v}}_d\frac{\partial(\hat{\mathbf{v}}_d \cdot \mathbf{v}_d)}{\partial t}\frac{\tau}{T}H_{5,i}+\tau\frac{\partial (\hat{\mathbf{v}}_d \cdot \mathbf{v}_d)}{\partial t}\mathbf{v}_{d} v_{d}H_{6,i}
  \Bigg\}.
\end{align}
The $H$ functions are defined as,
\begin{align}\label{H1}
   H_{1,i}=&\frac{1}{3\pi^{2}}\Bigg\{m_i^{2}T \sum_{j=1}^{\infty}(-1)^{j+1}\cosh{\left(\frac{j\mu}{T}\right)}K_{2}\left(\frac{jm_i}{T}\right)-m_i^{3}\sum_{j=1}^{\infty}(-1)^{j+1}j\cosh{\left(\frac{j\mu}{T}\right)}K_{1}\left(\frac{jm_i}{T}\right)\nonumber\\
    &\qquad\quad+m_i^{3}\sum_{j=1}^{\infty}(-1)^{j+1}j\cosh{\left(\frac{j\mu}{T}\right)}K_{\mathcal{I},1}\left(\frac{jm_i}{T}\right)\Bigg\},\nonumber\\
H_{2,i}=&\frac{1}{10\pi^{2}T^2}\Bigg\{3m_i^{3}T^{2}\sum_{j=1}^{\infty}(-1)^{j+1}j\cosh{\left(\frac{j\mu}{T}\right)}K_{3}\left(\frac{jm_i}{T}\right)
-\frac{m_i^{5}}{4}\sum_{j=1}^{\infty}(-1)^{j+1}j^{3}\cosh{\left(\frac{j\mu}{T}\right)}K_{3}\left(\frac{jm_i}{T}\right)\nonumber\\
 &\qquad\qquad\quad+\frac{5m_i^{5}}{4}\sum_{j=1}^{\infty}(-1)^{j+1}j^{3}\cosh{\left(\frac{j\mu}{T}\right)}K_{1}\left(\frac{jm_i}{T}\right)-m_i^{5}\sum_{j=1}^{\infty}(-1)^{j+1}j^{3}\cosh{\left(\frac{j\mu}{T}\right)}K_{\mathcal{I},1}\left(\frac{jm_i}{T}\right)\Bigg\},\nonumber\\
 H_{3,i}=&\frac{v_{d}}{T}\frac{1}{10\pi^{2}}\Bigg\{3m_i^{3}T^{2}\sum_{j=1}^{\infty}(-1)^{j+1}j\sinh{\left(\frac{j\mu}{T}\right)}K_{3}\left(\frac{jm_i}{T}\right)
 -\frac{m_i^{5}}{4}\sum_{j=1}^{\infty}(-1)^{j+1}j^{3}\sinh{\left(\frac{j\mu}{T}\right)}K_{3}\left(\frac{jm_i}{T}\right)\nonumber\\
 &\qquad\qquad\quad+\frac{5m_i^{5}}{4}\sum_{j=1}^{\infty}(-1)^{j+1}j^{3}\sinh{\left(\frac{j\mu}{T}\right)}K_{1}\left(\frac{jm_i}{T}\right)
 -m_i^{5}\sum_{j=1}^{\infty}(-1)^{j+1}j^{3}\sinh{\left(\frac{j\mu}{T}\right)}K_{\mathcal{I},1}\left(\frac{jm_i}{T}\right)\Bigg\},\nonumber\\
 H_{4,i}=&\frac{ m_i^{2}}{2\pi^{2}}\sum_{j=1}^{\infty}(-1)^{j+1}\cosh\left({\frac{j\mu}{T}}\right)K_{2}\left(\frac{jm_i}{T}\right),\nonumber\\
 H_{5,i}=&\frac{m_i^{2}T^{2}}{\pi^{2}}\sum_{j=1}^{\infty}{(-1)^{j+1}}\sinh\left({\frac{j\mu}{T}}\right)K_{2}\left(\frac{jm_i}{T}\right),\nonumber\\
H_{6,i}=&\frac{3m_i^{3}}{2\pi^{2}}\sum_{j=1}^{\infty}(-1)^{j+1}\sinh\left({\frac{j\mu}{T}}\right)K_{3}\left(\frac{jm_i}{T}\right).
\end{align}
The component $\mathbf{J}^{(2)}$ is the dissipative contributions to the current density in a drifting medium. This term arises from non-equilibrium effects associated with spatial and temporal variations in the plasma parameters. In particular, such contributions appear when gradients in thermodynamic quantities or flow fields perturb the local equilibrium distribution. These new types of contributions arising from plasma inhomogeneities in a relativistic plasma, in the massless limit, were first reported in Ref.~\cite{Gorbar:2016qfh}. In a drifting plasma, the presence of a background drift velocity modifies the equilibrium distribution, and any space-time variation of this drift flow leads to additional corrections to the current density. Consequently, $\mathbf{J}^{(2)}$ get contributions from gradients of the drift velocity itself, which characterize the inhomogeneity of the flow. In addition, variations in the chemical potential also contribute to this dissipative current.

Similarly, the charge density of the drifting plasma at the equilibrium can be estimated from Eq.~(\ref{current}) as,
\begin{align}
   J^0\equiv \rho = \sum_{i} g_{i} q_i \int \frac{d^{3}p}{(2\pi)^{3}}f_{\mathrm{eq},i}^{(v_d)}(t,x,p),
\end{align}
where $f_{\mathrm{eq}\,\, i}^{(v_d)}$ is defined in Eq.~(\ref{Particle Distribution Function}). Solving the integral, we obtain, 
\begin{align}\label{n_total}
    \rho&=\sum_{i} g_iq_i\Bigg\{\frac{m_i^{2}T}{\pi^{2}} \sum_{j=1}^{\infty} \frac{(-1)^{j+1}}{j}\sinh\left(\frac{j\mu}{T}\right) K_{2}\left(\frac{jm_i}{T}\right)
    +\frac{{v}_d^{2} m_i^{3}}{2\pi^{2}} \sum_{j=1}^{\infty} (-1)^{j+1} \sinh\left(\frac{j\mu}{T}\right) K_{3}\left(\frac{jm_i}{T}\right)\Bigg\}.
\end{align}
The first term in Eq.~(\ref{n_total}) corresponds to the charge density of the plasma in the absence of the drift, while the second term arises as a correction due to the finite drift velocity of the medium. Although this correction is typically sub-leading for small drift velocities, it becomes relevant when the collective motion of the plasma is significant. Note that both terms are proportional to $\sinh(j\mu/T)$, indicating that the charge density is explicitly depends on the quark chemical potential and net charge density of the plasma vanishes in the limit $\mu \rightarrow 0$.
\subsection*{Case II: Polarization Drift}
In the presence of a time-dependent electric field $\mathbf{E}(t)$, an additional drift component arises due to the temporal variation of the field. This effect is known as the polarization drift \cite{chen2016introduction}.
The total drift velocity in such a situation can be written as,
\begin{align}\label{v_total}
     \mathbf{v}_d\pm \mathbf{v}_{p}
            = \,\frac{\mathbf{E} \times \mathbf{B}}{B^{2}} 
            \pm \frac{1}{\omega_{c} B}\,\frac{d\mathbf{E}}{dt},
\end{align}
\begin{figure}
    \centering
    \includegraphics[width=8cm,height=9cm]{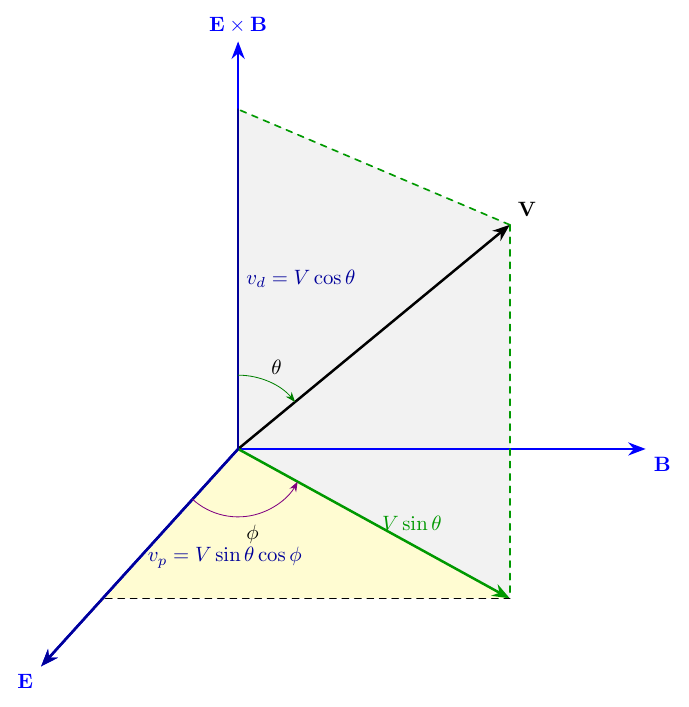}
    \caption{Geometry of the drift configuration in crossed electric and magnetic fields. 
The drift direction is given by $\mathbf{E}\times\mathbf{B}$. 
The particle velocity $\mathbf{V}$ is decomposed into $v_d=V\cos\theta$ along the drift direction and a perpendicular component $v_\perp=V\sin\theta$. 
The projection $v_p=V\sin\theta\cos\phi$ corresponds to the polarization drift and contributes to the polarization current.}
    \label{EB}
\end{figure}
where first term is the standard $\mathbf{E}\times \mathbf{B}$ drift ${\bf v}_d$ and the second term represents the polarization drift ${\bf v}_{p}$, which appears due to the time-varying nature of the electric field. 
The quantity $\omega_{c}$ denotes the \textit{cyclotron frequency}, $\omega_{c} = \frac{|q|B}{\epsilon}$. Unlike the standard $\mathbf{E}\times \mathbf{B}$ drift, the polarization drift depends on the charge of the particle. Hence, the drift will be in opposite directions for particle and antiparticle. In this case, the current density can be defined as,
\begin{align}\label{J Total vp}
\mathbf{J}^{(\text{Total})}
&= \sum_i g_i q_i 
\Bigg\{
\int \frac{d^3p}{(2\pi)^3}\,\mathbf{V}\, f^{(v_d,v_p)}_{i }
+ \int \frac{d^3p}{(2\pi)^3}\,\frac{\tau D^{(v_d,v_p)}_{i }}{T}\,
\mathbf{V}\,q_i\,\frac{(\mathbf{E}\!\cdot\!\mathbf{B})(\mathbf{V}\!\cdot\!\mathbf{B})}{B^2}
\nonumber\\[4pt]
&\qquad
- \int \frac{d^3p}{(2\pi)^3}\,\mathbf{V}
\left(\mathbf{V}\!\cdot\!\frac{\partial \mu}{\partial \mathbf{X}}\right)
\frac{\tau D^{(v_d,v_p)}_{i }}{T}
- \int \frac{d^3p}{(2\pi)^3}\,\mathbf{V}
\left(\mathbf{V}\!\cdot\!\frac{\partial (\mathbf{p}\!\cdot\!\mathbf{v}_d)}{\partial \mathbf{X}}\right)
\frac{\tau D^{(v_d,v_p)}_{i}}{T}
\nonumber\\[4pt]
&\qquad
-\int \frac{d^3p}{(2\pi)^3}\,\mathbf{V}\,
\frac{\partial \mu}{\partial t}\,
\frac{\tau D^{(v_d,v_p)}_{i }}{T}
- \int \frac{d^3p}{(2\pi)^3}\,\mathbf{V}\,
\frac{\partial (\mathbf{p}\!\cdot\!\mathbf{v}_d)}{\partial t}\,
\frac{\tau D^{(v_d,v_p)}_{i }}{T}
\Bigg\}\\
&={J}_d \hat{\mathbf{v}}_d+{J}_p \hat{\mathbf{v}}_p.
 \end{align}
The presence of polarization drift, driven by the time-dependent electric field, modifies the distribution function by introducing an additional drift velocity component. This component arises from the temporal change in the electric field, which causes a shift in the guiding center motion of particles. We have
\begin{equation}\label{Distribution Function with polarization drift}
f^{(v_d,v_p)}(t, \mathbf{x}, \mathbf{p}) = 
\frac{1}{\exp\left[ \beta \left( \epsilon - \mathbf{p} \cdot (\mathbf{v}_d+\mathbf{v}_p) - \mu(t, \mathbf{x}) \right) \right] + 1}.
\end{equation}
We expand the modified distribution function in Eq.~\eqref{Distribution Function with polarization drift} with respect to $v_p$, following the procedure described in Appendix~\ref{Expansion of the Drift-Modified Distribution Function}. We obtain,
\begin{equation}
    f^{(v_{d},v_p)}(t,x,p)=f^{(v_d)}+D^{(v_d)}\frac{pv_p\sin\theta\cos \phi}{T}+D^{(v_d)}(1-2f^{(v_d)})\frac{p^{2}{v}_p^{2}\sin^{2}\theta \cos^2 \phi}{2T^{2}}+...
\end{equation}
with
\begin{align}
\begin{split}
    D^{(v_{d},v_{p})}=&D^{(v_d)}+D^{(v_d)}\left(1-2f^{(v_d)}\right)\frac{p{v}_p\sin\theta \cos\phi}{T}+D^{(v_d)}\left(1-6f^{(v_d)}+6f^{(v_d)2}\right)
    \frac{p^{2}{v}_p^{2}\sin^{2}\theta \cos^{2}\phi}{2T^{2}}+...
    \end{split}
\end{align}
In the present perturbative treatment, the expansion is organized in terms of the small drift-induced corrections entering the distribution function. Consequently, contributions linear in $v_p$ are retained, while mixed terms proportional to $v_d v_p$ and $v^2_p$ are consistently neglected. We have,
\begin{equation}\label{feqvp}
    f^{(v_{d},v_p)}(t,x,p)=f^{(v_d)}+D_{eq}\frac{pv_p\sin\theta\cos \phi}{T},
\end{equation}
where,
\begin{equation}\label{Deqvp}
    D^{(v_{d},v_{p})}(t,x,p)=D^{(v_d)}+D_{eq}(1-2f_{eq})\frac{p{v}_p\sin\theta \cos\phi}{T}.
\end{equation}
The total current obtained in this case can be decomposed into contributions along the drift and polarization directions. The component parallel to the drift velocity corresponds to the usual drift current, while the perpendicular component arises from the polarization drift induced by the time-dependent electric field. The drift current takes the form as,
\begin{align}
\mathbf{J}_d
&= \sum_i g_i q_i 
\Bigg\{
\int \frac{d^3p}{(2\pi)^3}\,{V}\cos\theta\, \hat{\mathbf{v}}_d\left(f^{(v_d)}_i+D_{\mathrm{eq},i}\frac{pv_p\sin\theta\cos \phi}{T}\right)\nonumber\\
&+ \int \frac{d^3p}{(2\pi)^3}\,\frac{\tau}{T}\,
{V} \cos\theta\,\hat{\mathbf{v}}_d q_i\,\frac{(\mathbf{E}\!\cdot\!\mathbf{B})(\mathbf{V}\!\cdot\!\mathbf{B})}{B^2}
\left(D^{(v_d)}_i+D_{\mathrm{eq},i}(1-2f_{\mathrm{eq},i})\frac{p{v}_p\sin\theta \cos\phi}{T}\right)\nonumber\\
&
- \int \frac{d^3p}{(2\pi)^3}\,({V} \cos\theta\,)^2\hat{\mathbf{v}}_d
\left(\hat{\mathbf{v}}_d \!\cdot\!\frac{\partial \mu}{\partial \mathbf{X}}\right)
\frac{\tau}{T}\left(D^{(v_d)}_i+D_{\mathrm{eq},i}(1-2f_{\mathrm{eq},i})\frac{p{v}_p\sin\theta \cos\phi}{T}\right)\nonumber\\
&- \int \frac{d^3p}{(2\pi)^3}\,({V} \cos\theta\,)^2\hat{\mathbf{v}}_d
\left(\hat{\mathbf{v}}_d \!\cdot\!\frac{\partial (\mathbf{p}\!\cdot\!\mathbf{v}_d)}{\partial \mathbf{X}}\right)
\frac{\tau}{T}\left(D^{(v_d)}_i+D_{\mathrm{eq},i}(1-2f_{\mathrm{eq},i})\frac{p{v}_p\sin\theta \cos\phi}{T}\right)
\nonumber\\
&-\int \frac{d^3p}{(2\pi)^3}\,{V}\cos\theta\,\hat{\mathbf{v}}_d\,
\frac{\partial \mu}{\partial t}\,
\frac{\tau }{T}\left(D^{(v_d)}_i+D_{\mathrm{eq},i}(1-2f_{\mathrm{eq},i})\frac{p{v}_p\sin\theta \cos\phi}{T}\right)\nonumber\\
&- \int \frac{d^3p}{(2\pi)^3}\,{V}\cos\theta\,\hat{\mathbf{v}}_d
\frac{\partial (\mathbf{p}\!\cdot\!\mathbf{v}_d)}{\partial t}\,
\frac{\tau }{T}\left(D^{(v_d)}_i+D_{\mathrm{eq},i}(1-2f_{\mathrm{eq},i})\frac{p{v}_p\sin\theta \cos\phi}{T}\right) \Bigg\}.
\end{align}
Upon performing the momentum integration, it reproduces the same form obtained previously in Eq.~\eqref{J Total} in the limit of constant electric and magnetic fields, confirming the consistency of the formulation. Similarly, we can define the polarization current from Eq.~(\ref{J Total vp}) as,
\begin{align}
\mathbf{J}_p
&= \sum_i g_i q_i 
\Bigg\{
\int \frac{d^3p}{(2\pi)^3}\,{V}\sin\theta \cos\phi\, \hat{\mathbf{v}}_p\left(f^{(v_d)}_i+D_{\mathrm{eq},i}\frac{pv_p\sin\theta\cos \phi}{T}\right)\nonumber\\
&+ \int \frac{d^3p}{(2\pi)^3}\,\frac{\tau}{T}\,
{V}\sin\theta \cos\phi\,\hat{\mathbf{v}}_p q_i\,\frac{(\mathbf{E}\!\cdot\!\mathbf{B})(\mathbf{V}\!\cdot\!\mathbf{B})}{B^2}
\left(D^{(v_d)}_i+D_{\mathrm{eq},i}(1-2f_{\mathrm{eq},i})\frac{p{v}_p\sin\theta \cos\phi}{T}\right)\nonumber\\
&
- \int \frac{d^3p}{(2\pi)^3}\,({V}\sin\theta \cos\phi\,)^2 \hat{\mathbf{v}}_p
\left(\hat{\mathbf{v}}_p \!\cdot\!\frac{\partial \mu}{\partial \mathbf{X}}\right)
\frac{\tau}{T}\left(D^{(v_d)}_i+D_{\mathrm{eq},i}(1-2f_{\mathrm{eq},i})\frac{p{v}_p\sin\theta \cos\phi}{T}\right)\nonumber\\
&- \int \frac{d^3p}{(2\pi)^3}\,({V}\sin\theta \cos\phi\,)^2\hat{\mathbf{v}}_p
\left(\hat{\mathbf{v}}_p\!\cdot\!\frac{\partial (\mathbf{p}\!\cdot\!\mathbf{v}_p)}{\partial \mathbf{X}}\right)
\frac{\tau}{T}\left(D^{(v_d)}_i+D_{\mathrm{eq},i}(1-2f_{\mathrm{eq},i})\frac{p{v}_p\sin\theta \cos\phi}{T}\right)
\nonumber\\
&-\int \frac{d^3p}{(2\pi)^3}\,{V}\sin\theta \cos\phi\,\hat{\mathbf{v}}_p\,
\frac{\partial \mu}{\partial t}\,
\frac{\tau }{T}\left(D^{(v_d)}_i+D_{\mathrm{eq},i}(1-2f_{\mathrm{eq},i})\frac{p{v}_p\sin\theta \cos\phi}{T}\right)\nonumber\\
&- \int \frac{d^3p}{(2\pi)^3}\,{V}\sin\theta \cos\phi\,\hat{\mathbf{v}}_p\,
\frac{\partial (\mathbf{p}\!\cdot\!\mathbf{\mathbf{v}}_p)}{\partial t}\,
\frac{\tau }{T}
\left(D^{(v_d)}_i+D_{\mathrm{eq},i}(1-2f_{\mathrm{eq},i})\frac{p{v}_p\sin\theta \cos\phi}{T}\right) \Bigg\}.
\end{align}
Solving the integral, we obtain the polarization current in a drifting plasma as follows,
\begin{equation}\label{JP}
    \begin{split}
        \mathbf{J}_p&=\sum_i g_i\frac{m_i^3 T}{\pi^2 B_0^2}\frac{d\mathbf{E}}{dt} \sum_{j=1}^{\infty}\frac{(-1)^{j+1}}{j}\cosh\left(\frac{j\mu}{T}\right)K_3\left(\frac{jm_i}{T}\right)+\sum g_i q_i^2\Bigg\{\frac{\tau}{T}\frac{(\mathbf{E}\cdot\mathbf{B})(\mathbf{B}\cdot \hat{\mathbf{v}}_p)}{B^{2}}\left(H_{1,i}+v_d^2\frac{H_{2,i}}{3}\right)\Bigg\}\hat{\mathbf{v}}_p\\
        &-\sum_i g_i q_i\Big(\hat{\mathbf{v}}_p\cdot \frac{\partial \mu}{\partial \mathbf{X}}\Big)\frac{\tau}{T}\left(H_{1,i}+v_d^2\frac{H_{2,i}}{3}\right) \hat{\mathbf{v}}_p-\sum_i g_i\frac{m_i^3\tau}{\pi^2B^2}\frac{d\mathbf{E}}{dt}\frac{\partial\mu}{\partial t}\sum_{j=1}^{\infty}(-1)^{j+1}\cosh\left(\frac
        {j\mu}{T}\right)K_3\left(\frac{jm_i}{T}\right) \\
        &-\sum_i g_i \frac{d^2 \mathbf{E}}{dt^2}\frac{m_i^{3}\tau }{\pi^2 B_0^2}\Bigg[T\sum_{j=1}^{\infty}\frac{(-1)^{j+1}}{j}\cosh\left(\frac{j\mu}{T}\right)K_3\left(\frac{jm_i}{T}\right)- \frac{v_d^2 m_i}{2}\sum_{j=1}^{\infty} (-1)^{j+1}\cosh\left(\frac{j\mu}{T}\right)K_4\left(\frac{jm_i}{T}\right)\Bigg].
\end{split}
\end{equation}
 The term proportional to $d\mathbf{E}/dt$ gives the leading polarization-drift current generated by the time variation of the electric field. 
The field-dependent terms involving $(\mathbf{E}\!\cdot\!\mathbf{B})(\mathbf{B}\!\cdot\!\hat{\mathbf v}_p)/B^2$ arise from the anisotropic coupling of the electric and magnetic fields in the drift-modified medium. 
The terms containing $\partial\mu/\partial X$ and $\partial\mu/\partial t$ describe the response of the polarization current to spatial and temporal variations of the local chemical potential. 
The contribution proportional to $d^2\mathbf{E}/dt^2$ corresponds to a higher-order dynamical correction associated with the time dependence of the polarization response itself.
\section{Quantitative Implementation of the Results in QGP Medium}\label{section III}
While applying this framework to the QGP, it is important to note that the medium contains electrically neutral gluons, which interact with the charged quarks and provide an effective drag. As emphasized in an earlier study~\cite{Gorbar:2016qfh}, this can hinder the development of a fully collective drift of the plasma. Consequently, the boosted drift-equilibrium distribution should be interpreted as an effective or instantaneous reference state for the charged sector rather than an exact equilibrium state of the full medium. We consider the QGP medium in the regime where interactions between neutral constituents (gluons) and charged particles (quarks and antiquarks) are relatively weak such that the FD distribution for the charged species is modified, as they acquire a collective drift under the influence of electromagnetic fields present in the medium. In this approach, the effects of strong interactions are incorporated phenomenologically through the relaxation-time parameter. In a hot and dense QCD plasma, the relaxation time is governed by microscopic scattering processes among the charge carriers and is therefore sensitive to the interaction strength of the medium. In the QGP, this interaction strength is determined by the running QCD coupling constant, which depends on the characteristic energy scale of the medium as well as on external parameters such as the magnetic field and chemical potential.
In the present analysis, we focus on the weak-field regime and neglect the effects of Landau level quantization on the phase space and scattering processes. Under these assumptions, the relaxation time for quarks of a given flavor can be estimated within perturbative QCD, where it is governed by elastic scattering processes and logarithmic screening effects. Consequently, the relaxation time $\tau$ becomes explicitly temperature dependent and can be expressed in terms of the running QCD coupling $\alpha_s$. The expression for relaxation time is given by \cite{Hosoya:1983xm},
\begin{align}
\tau(T) = 
\frac{1}{
5.1\, T\, \alpha_s^2 \log(1/\alpha_s)\, \left[1 + 0.12\, (2N_f + 1)\right]
}.
\end{align}
Here, $\alpha_s$ is the QCD running coupling constant \cite{KapustaLandshoff1989}
\begin{align}
\alpha_s(\Lambda^2, eB) = 
\frac{\alpha_s(\Lambda^2)}{
1 + b_1\, \alpha_s(\Lambda^2)\, 
\ln\!\left(\frac{\Lambda^2}{\Lambda^2 + eB}\right)
}.
\end{align}
Where,
\begin{align}
\alpha_s(\Lambda^2) = 
\frac{1}{
b_1\, \ln\!\left(\dfrac{\Lambda^2}{\Lambda^2_{\text{MS}}}\right)
}.
\end{align}
Here, $\Lambda=2\pi \sqrt{T^2 + \left(\frac{\mu}{\pi}\right)^2 }$ denotes the characteristic thermal scale of the medium.
This sets the natural renormalization scale for evaluating the running QCD coupling in a hot and dense plasma. 
The parameter $\Lambda_{\mathrm{MS}}$ denotes the QCD scale parameter defined in the modified minimal subtraction (MS) renormalization scheme. For the high-temperature QGP medium, we take $\Lambda_{\mathrm{MS}} = 0.176~\text{GeV}$ for the one-loop running coupling, consistent with lattice measurements \cite{Bazavov2012DeterminationO} and the Particle Data Group \cite{ParticleDataGroup:2012pjm}. The coefficient $b_1$ is the one-loop coefficient of the QCD beta function and determines the scale dependence of the running coupling $\alpha_s$. It depends on the number of colors $N_c$ and the number of active quark flavors $N_f$, and is given by
\begin{align}
b_1 = \frac{11N_c - 2N_f}{12\pi}.
\end{align}
For quantitative analysis of the results, we consider two scenarios of a homogeneous plasma. 
\subsubsection*{Case I: Constant electric and magnetic fields}
We consider the case in which the QGP is subjected to constant and mutually perpendicular electric and magnetic fields $\mathbf{E}=E_{0}\hat{{\bf y}},\,\,  \mathbf{B}=B_{0}\hat{{\bf z}}$. For this configuration, the corresponding drift velocity takes the form $\mathbf{v}_d=v_d\,\hat{{\bf x}}$, with $v_d=\frac{E_0}{B_0}$. For a homogeneous plasma, we obtain the charge density of the drifting QGP from Eq.~\eqref{n_total} as,
\begin{align}
    \rho
    =&{2N_cq_s\bigg\{\frac{m_s^{2}T}{\pi^{2}} \sum_{j=1}^{\infty} \frac{(-1)^{j+1}}{j} \sinh\left({\frac{j\mu}{T}}\right) K_{2}\left(\frac{jm_s}{T}\right)+\frac{{v}_d^{2} m_s^{3}}{2\pi^{2}} \sum_{j=1}^{\infty} (-1)^{j+1} \sinh\left(\frac{j\mu}{T}\right) K_{3}\left(\frac{jm}{T}\right)\Bigg\}}\nonumber\\
    &+\sum_{i=u,d}\frac{2N_cq_iT^3}{\pi^2}(1+{2v_d^2})\left[\text{Li}_3\left\{-\exp\left(-\frac{\mu}{T}\right)\right\}-\text{Li}_3\left\{-\exp\left(\frac{\mu}{T}\right)\right\}\right].
\end{align}
The first two term denotes the contribution from the strange quarks, and the last term corresponds to the up and down quark contributions to the charge density. As the $u$ and $d$ quarks are significantly lighter than the $s$ quark, we adopt the massless limit for those flavors. The massive-to-massless conversion of the thermodynamic integrals is discussed in detail in the Appendix~\ref{Mass-to-Massless Conversion}. Similarly, we estimate the current density from Eq.~(\ref{J Total}) for constant fields in a homogeneous plasma as,
\begin{align}\label{J_x case I}
      {J}_x=&2N_cq_s\frac{E_0}{B_0}\frac{m_s^{2}T}{\pi^{2}}\sum_{j=1}^{\infty}\frac{(-1)^{j+1}}{j}\sinh{\left(\frac{j\mu}{T}\right)}K_{2}\left(\frac{jm_s}{T}\right)\nonumber\\
        &+2N_c\sum_{i=u, d} q_i\frac{E_0}{B_0}\frac{T^3}{\pi^2}\left[\text{Li}_3\left\{-\exp\left(-\frac{\mu}{T}\right)\right\}-\text{Li}_3\left\{-\exp\left(\frac{\mu}{T}\right)\right\}\right].
\end{align}
This component represents the Hall current of the plasma in the direction perpendicular to both electric and magnetic field. Note that $J_x$ component is sensitive to the net charge density of the medium, which is governed by the quark chemical potential $\mu$. In the limit $\mu = 0$, the quarks and antiquarks contributions to the Hall current cancel each other. 
\subsubsection*{Case II: Temporal electric field}  
To incorporate polarization drift in the analysis, we consider a constant magnetic field $\mathbf{B}=B_{0}\hat{{\bf z}}$ and a time-dependent electric field $\mathbf{E}(t)=E_{0}e^{-\frac{t}{\tau_E}}\hat{{\bf y}}$, where $\tau_E$ is the characteristic time over which the electric field decreases\footnote{For numerical illustration, we set $\tau_E = 5~fm/c$, corresponding to a characteristic time scale of the electromagnetic field evolution.} Along with the $\mathbf{v}_d$, now we have the polarization drift due to the temporal change in the electric field as,
\begin{align}\label{Polarization Drift}
   \mathbf{v}_{p} =\mp \frac{\sqrt{p^2 +m^2}}{\mid q\mid B_0^2}\frac{E_0 e^{-\frac{t}{\tau_E}}}{\tau_E}\hat{{\bf y}}=\mp \frac{\sqrt{p^2 +m^2}}{\mid q\mid B_0^2}\frac{E(t)}{\tau_E}\hat{{\bf y}}
\end{align}
In this scenario, the current has two components.  The first component is directed along the $x$-axis, $i.e.$, along the drift velocity $\mathbf{v}_d$, and is denoted by ${J}_x$. 
The second component is directed along the $y$-axis, corresponding to the polarization drift, and is denoted by ${J}_y$. The component ${J}_x$ takes the form as follows,
\begin{align}\label{J_x case II}
         {J}_x
        =&2N_cq_s\frac{E(t)}{B_0}\frac{m_s^{2}T}{\pi^{2}}\sum_{j=1}^{\infty}\frac{(-1)^{j+1}}{j}\sinh{\left(\frac{j\mu}{T}\right)}K_{2}\left(\frac{jm_s}{T}\right)\nonumber\\
        &+2N_c\sum_{i=u, d} q_i\frac{E(t)}{B_0}\frac{T^3}{\pi^2}\left[\text{Li}_3\left\{-\exp\left(-\frac{\mu}{T}\right)\right\}-\text{Li}_3\left\{-\exp\left(\frac{\mu}{T}\right)\right\}\right]\nonumber\\
         &+2N_cq_s\frac{E(t)}{\tau_EB_0}\frac{m_s^{2}T\tau}{\pi^{2}}\sum_{j=1}^{\infty}(-1)^{j+1}\sinh\left({\frac{j\mu}{T}}\right)K_{2}\left(\frac{jm_s}{T}\right)\nonumber\\
         &+2N_c\sum_{i= u, d}q_i\frac{E (t)}{\tau_E B_0}\frac{\tau T^3}{\pi^2}\left[\text{Li}_2\left\{-\exp\left(-\frac{\mu}{T}\right)\right\}-\text{Li}_2\left\{-\exp\left(\frac{\mu}{T}\right)\right\}\right].
\end{align}
In comparison to the {\it case I} (as described in Eq.~(\ref{J_x case I})), $J_x$ is modified due the time dependence of the electric field in the medium. The last two terms of Eq.~(\ref{J_x case II}) are the additional corrections solely due to the temporal variation of the electric field in the medium. The correction term depends on the collision dynamics of the medium through the thermal relaxation time $\tau$. In this specific choice of field configuration, the choice of $\tau_E$ has an explicit dependence on the correction terms. Similar to Eq.~(\ref{J_x case I}), Eq.~(\ref{J_x case II}) vanishes at $\mu=0$.  
Similarly, we obtain ${J}_y$ component as,
\begin{align}
{J}_y
&= \frac{2N_c}{\pi^2}\frac{E(t)}{\tau_E B_0^2}\Bigg[{-m_s^3 T \sum_{j=1}^{\infty}\frac{(-1)^{j+1}}{j}\cosh\left(\frac{j\mu}{T}\right)K_3\left(\frac{jm_s}{T}\right)}\nonumber\\
&\qquad\qquad\qquad\quad+8T^4\left[\text{Li}_4\left\{-\exp\left(-\frac{\mu}{T}\right)\right\}+\text{Li}_4\left\{-\exp\left(\frac{\mu}{T}\right)\right\}\right]\Bigg].
\end{align}
In a QGP medium, both quarks and antiquarks act as charge carriers and contribute to the $J_y$ component. Unlike the $J_x$ component, $J_y$ does not vanish in the limit $\mu = 0$. Even at vanishing chemical potential, the medium contains thermally excited quark-antiquark pairs and these charged particles respond to the time-dependent electric field and give rise to a finite current. 

\begin{figure}
  	\begin{center}
  		{\centering
  		\subfloat[]{\includegraphics[width=8.1cm,height=6cm]{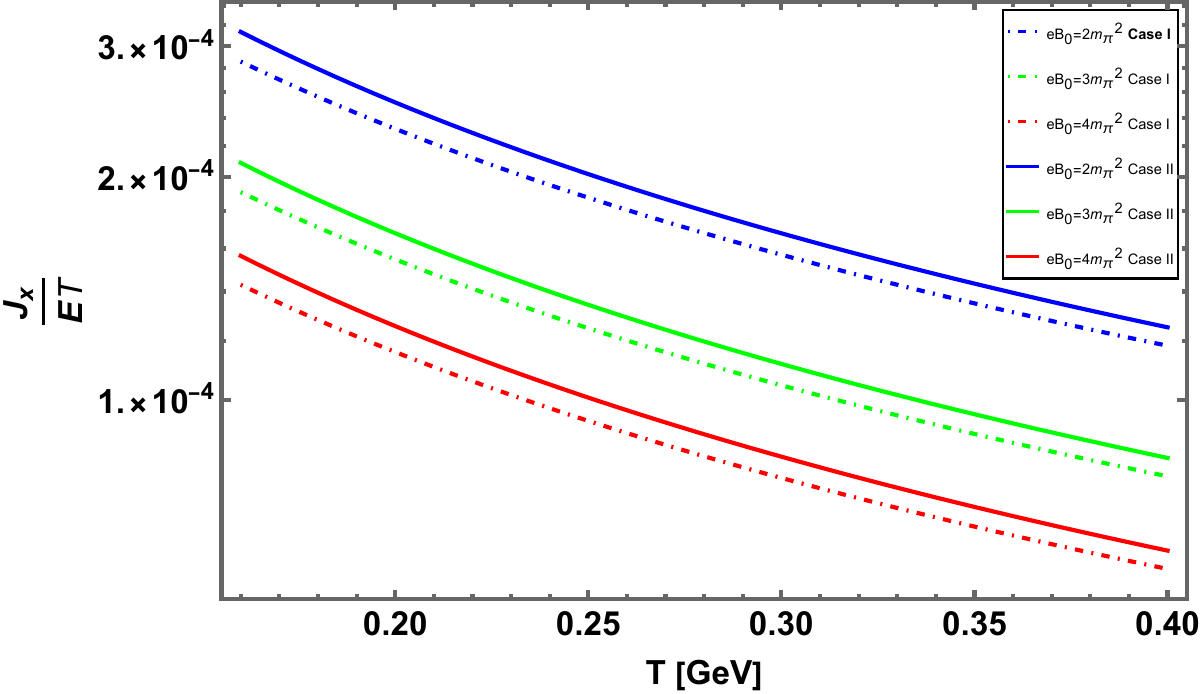}\label{JdT}} \hspace {0.0cm} 
  			\subfloat[]{\includegraphics[width=8.1cm,height=6cm]{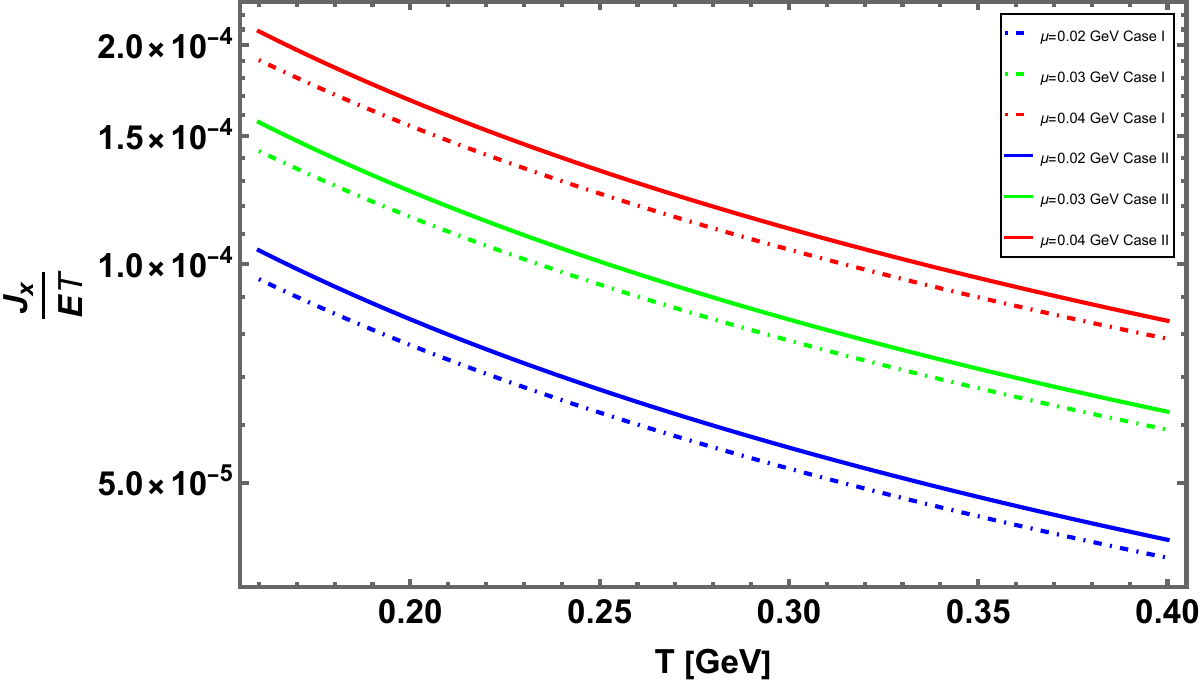} \label{JdT1}} 
  			\caption{(a) Variation of $\frac{J_x}{ET}$ of {\it case I} and {\it case II}, with temperature at fixed chemical potential, $\mu=0.04$ GeV. (b) Variation of  $\frac{J_x}{ET}$ of {\it case I} and {\it case II}, with temperature at a fixed magnetic field, $eB_0=3m_{\pi}^{2}$.}}
  	\end{center}
  \end{figure}

The variation of the dimensionless quantity $\frac{J_x}{ET}$ with temperature is shown in Fig.~\ref{JdT} for both cases. The dashed curves denotes {\it case I} of constant electromagnetic fields, and the solid lines represents the {\it case II} with temporal electric field. The current component $J_x$ arises due to the ${\mathbf E}\times {\mathbf B}$ drift in the QGP. As the temperature increases, the magnitude of the $J_x/ET$ gradually decreases over the considered temperature range for both the cases. This behavior can be understood from the fact that increasing temperature enhances the thermal motion of the charge carriers in the plasma, which reduces the relative contribution of the drift motion induced by the external electromagnetic fields. Consequently, the drift component of the current becomes less significant at higher temperatures. The $J_x$ component in the {\it case I}, given in Eq.~(\ref{J_x case I}) is modified when the time-dependent behavior of the electric field in the medium is considered, as expressed in Eq.~(\ref{J_x case II}). We observe that the leading-order contribution to $J_x$ originates from the drift-modified equilibrium distribution of charged particle in the medium, whereas the non-equilibrium contribution appears only as a sub-leading order correction. We have also study the impact of the strength of the magnetic field in the $J_x$ component of the current. The sensitivity of the ${\mathbf E}\times {\mathbf B}$ drift current on the chemical potential is depicted in Fig.~\ref{JdT1}. It is seen that $\frac{J_x}{ET}$ increases with $\mu$. This behavior can be understood from the fact that a larger chemical potential corresponds to a higher charge density in the medium. As the number of charged particles increases, the response of the system to the external electromagnetic fields becomes stronger.
\begin{figure}
  	\begin{center}
  		{\centering
  		\subfloat[]{\includegraphics[width=8.1cm,height=6cm]{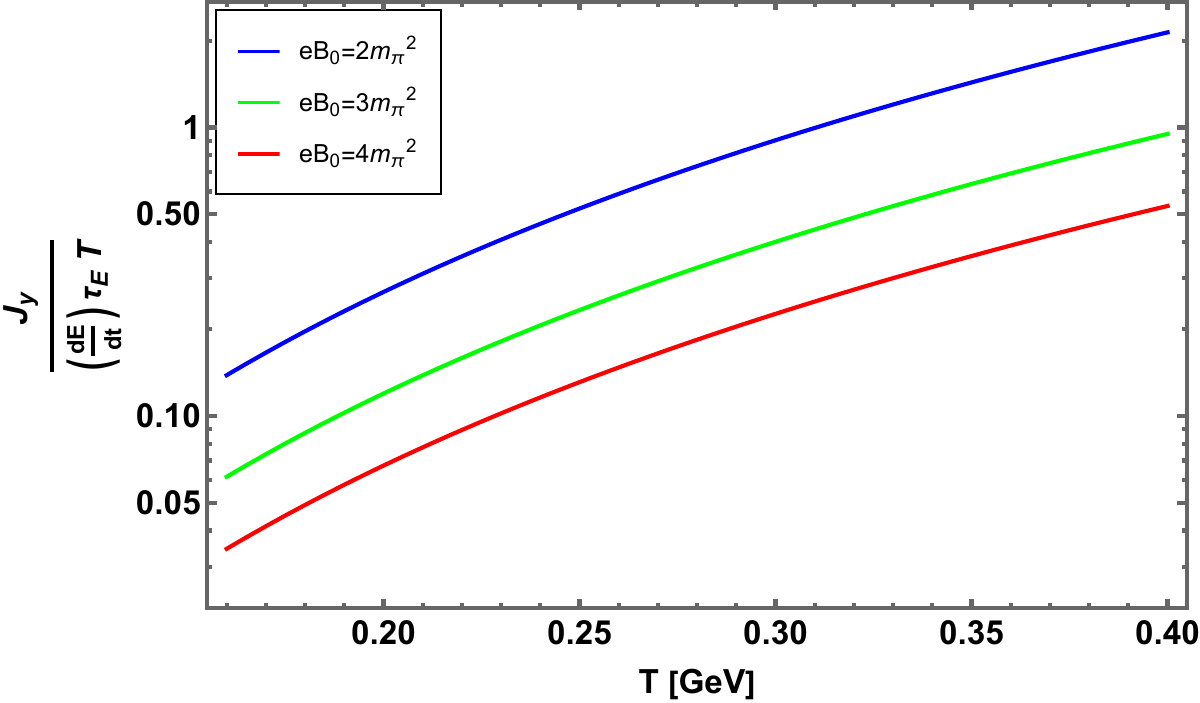}\label{JpT}} \hspace {0.0cm}
  			\subfloat[]{\includegraphics[width=8.1cm,height=6cm]{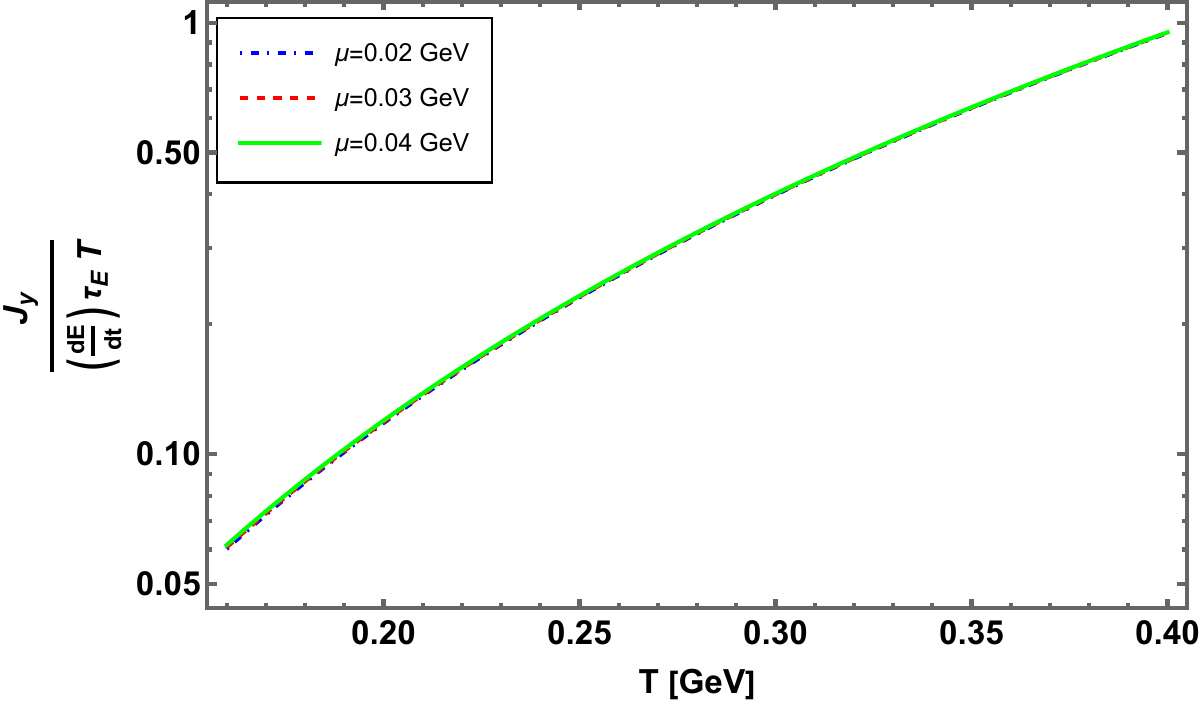} \label{JpT1}} 
  			\caption{(a) Variation of $\frac{J_y}{{\frac{dE}{dt}\tau_E T}}$ with temperature at fixed chemical potential, $\mu=0.04$ GeV (b) Variation of  $\frac{J_y}{{\frac{dE}{dt}\tau_E T}}$ with temperature at fixed magnetic field, $eB_0=3m_{\pi}^2$.}
  		}
  	\end{center}
  \end{figure}
  
Figure~\ref{JpT} shows the variation of the dimensionless quantity $\frac{J_y}{\big(\frac{dE}{dt}\big)\tau_E T}$ with temperature at different magnetic fields. In contrast to the ${\mathbf E}\times {\mathbf B}$ drift current, the polarization current exhibits a comparatively larger magnitude throughout the temperature range considered. This behavior can be understood from the relation $J_y/J_x \sim (m/qB)(1/E)(dE/dt)$, which indicates that the polarization current is enhanced when the electric field varies rapidly in time. The polarization current is directly driven by the temporal variation of the electric field.  In rapidly evolving relativistic plasmas, such as the QGP produced in heavy-ion collisions, electromagnetic fields vary significantly with time~\cite{Tuchin:2013apa,
McLerran:2013hla, Shovkovy:2022bnd}. As a result, the polarization drift develops as an immediate response to the changing electric field, while the drift current builds up only after the system relaxes through particle collisions.  We showed the dependence of the polarization current on the chemical potential in Fig.~\ref{JpT1}. It is observed that the impact of the chemical potential is negligible on the polarization current. This is because of the fact that the temperature is the dominant energy scale in the system in comparison with the chemical potential. Moreover, the $J_y$ component will not vanish when $\mu=0$. This behavior is quite opposite to that of the Hall current.

\section{Summary}\label{summary}
In this work, we investigated electric charge transport in a relativistic drifting plasma. Specifically, we analyzed two types of drift motion, the standard $\mathbf{E}\times\mathbf{B}$ drift, which arises from the presence of transverse electromagnetic fields, and the polarization drift, which occurs due to the time variation of the electric field in the medium. The $\mathbf{E}\times\mathbf{B}$ drift leads to the generation of a Hall current, while the polarization drift gives rise to a polarization current in the plasma. We obtained the general form of the current density in the drifting medium in an inhomogeneous plasma. To that end, we first estimated the drift-modified distribution function of the charged particles in the medium. The relativistic Boltzmann equation is solved within the RTA to obtain the non-equilibrium corrections to the distribution function, and consequently, to estimate the resulting non-equilibrium contributions to the current density. 

For quantitative implementation of the results, we consider the QGP medium in the regime where interaction of neutral particles (gluons) with charged particles (quarks/antiquarks) is relatively weak. In this scenario, the standard FD distribution will be modified for the charged particles as they experience a drift due to the presence of electromagnetic fields in the medium. In this analysis, we considered the massless limit for the up and down quarks, while the finite mass of the strange quark is retained. We evaluated the resulting current densities and study their temperature dependence on the magnetic field and chemical potential. This allows us to illustrate how conventional drift and polarization drift jointly influence the charge transport  of the QGP. Our results show that, in relativistic plasmas with time-dependent electromagnetic fields, the polarization current can become significantly larger than the conventional drift current. This behavior is closely related to the rapid temporal variation of electromagnetic fields in such systems and underscores the important role of polarization effects in shaping the electromagnetic response of the medium. We also observe that a similar qualitative dominance of polarization-type contributions over dissipative drift currents has been reported in earlier studies of relativistic plasma transport. In particular, the electromagnetic response of the QGP is known to deviate from the conventional Ohmic behavior on time scales shorter than the relaxation time~\cite{Shovkovy:2022bnd,Wang:2021oqq}. Furthermore, kinetic theory analyses of relativistic chiral plasmas reveal the presence of non-dissipative current components, such as magnetization and anomalous contributions~\cite{Gorbar:2016sey}.

The framework developed in this work offers a systematic method for studying drift-induced charge transport in relativistic plasmas under both static and time-dependent electromagnetic fields. The transport properties, however, depend on the specific choice of electromagnetic fields and their spacetime structure. While the formalism presented here is quite general, for realistic description, one may have to couple the current density equations derived from the kinetic theory with Maxwell’s equations that are relevant for the study of relativistic plasmas with inhomogeneities, which is beyond the scope of the present analysis. Another interesting aspect is to study the linear and nonlinear charge transport in a relativistic plasmas with a chiral asymmetry. We intend to explore these aspects in the near future.
\section{Acknowledgments}
A.S. acknowledges Indian Institute of Technology Bombay for the Institute Postdoctoral Fellowship. M.K. would like to acknowledge the Department of Science and Technology (DST), Govt. of India, for the INSPIRE-Faculty award (DST/INSPIRE/04/2024/001794), and the Faculty Research Scheme (FRS project number: MISC 0240) at IIT (ISM) Dhanbad. B.P. thanks CSIR-HRDG, Govt of India, for financial support received through Grant No. 03WS(003)/2023-24/EMR-II/ASPIRE. S. D. acknowledges the SERB Power Fellowship, SPF/2022/000014 for the support on this work. 

\appendix

\section{Drift Corrections to the Equilibrium Distribution}\label{Expansion of the Drift-Modified Distribution Function}
For many of the integrals appearing in the calculation of charge and current densities, the exact forms of the drift-modified distribution function $f_{eq}^{(v_{d})}$ and its derivative $D_{eq}^{(v_{d})}$ lead to expressions that are analytically complicated.  
 The relevant expansion parameter in the present case is the dimensionless drift correction $\frac{\mathbf{p}\cdot\mathbf{v}_d}{T},$
which, when the drift velocity $\mathbf{v}_d$ is chosen along the polar axis, reduces to $p v_d\cos\theta/T$. 
Since this parameter is small, the drift-modified distribution function can be expanded in a Taylor series in powers of $\mathbf{p}\cdot\mathbf{v}_d/T$.
This controlled expansion separates the equilibrium contribution from the drift-induced anisotropic corrections and allows the resulting momentum integrals to be evaluated a in closed form.  
In this appendix, we provide the explicit series expansions for $f_{eq}^{(v_{d})}$ and $D_{eq}^{(v_{d})}$ used throughout the main text, together with their higher-order drift corrections \cite{Dayi:2017xrr}.  
These expressions form the basis for obtaining the compact analytic forms of the number and current densities discussed in the main sections of the paper. We have
\begin{align}
    f_\mathrm{eq}^{(v_{d})}=f_\mathrm{eq}+D_{eq}\frac{p{v}_d\cos\theta}{T}+D_\mathrm{eq}(1-2f_\mathrm{eq})\frac{p^{2}{v}_d^{2}\cos^{2}\theta}{2T^{2}}+...
\end{align}
\begin{align}
    D_\mathrm{eq}^{(v_{d})}=D_\mathrm{eq}+D_\mathrm{eq}\left(1-2f_\mathrm{eq}\right)\frac{p{v}_d\cos\theta}{T}
    +D_\mathrm{eq}\left(1-6f_\mathrm{eq}+6f_\mathrm{eq}^{2}\right)\frac{p^{2}{v}_d^{2}\cos^{2}\theta}{2T^{2}}+...
\end{align}
Where, 
\begin{equation}
    f_\mathrm{eq}=\frac{1}{\exp\left[ \beta \left( \epsilon  - \mu \right) \right] + 1}.
\end{equation}
And, $D_\mathrm{eq}=f_\mathrm{eq}(1-f_\mathrm{eq})$.
\section{Fugacity Expansion of the Equilibrium Distribution Function}\label{Fugacity Expansion of the Fermi–Dirac Distribution}
In this appendix, we present the series expansions of the equilibrium distribution function and its derivatives that are used throughout the analytical calculations. 
The equilibrium distribution $f_{eq}$ and the combinations involving its derivative $D_{eq}$ admit rapidly convergent series representations in powers of $e^{-z}$, where $z=\beta(\epsilon-\mu)$. 
Such representations are particularly convenient because, after performing the momentum integrations, each exponential term naturally generates modified Bessel functions with argument $jm/T$. 
Consequently, these expansions provide the starting point for deriving the compact analytic expressions for the number and current densities in terms of modified Bessel functions. 
The relevant series representations are given below.
\begin{align}
  f_{eq}=\frac{1}{1+e^{z}}=\sum_{j=1}^{\infty}(-1)^{j+1}e^{-jz}
\end{align}
\begin{align}
   D_{eq}= \frac{e^{z}}{(1+e^{z})^{2}}=\sum_{j=1}^{\infty}(-1)^{j+1}je^{-jz} 
\end{align}
\begin{align}
   D_{eq}(1-2f_{eq})=\frac{e^{z}}{(1+e^{z})^{2}}\left(1-\frac{2}{(e^{z}+1)}\right)=\sum_{j=1}^{\infty}(-1)^{j+1}j^{2}e^{-jz}  
\end{align}
\begin{align}
   D_{eq}(1-6f_{eq}+6f_{eq}^{2})=\frac{e^{z}}{(1+e^{z})^{2}}\left(1-\frac{6}{(e^{z}+1)}+\frac{6}{(e^{z}+1)^{2}}\right)=\sum_{j=1}^{\infty}(-1)^{j+1}j^{3}e^{-jz}  
\end{align}
\section{Momentum Integrals and Bessel Function Structure}\label{General Momentum Integrals and Bessel Function Structure}
The integrals required in our analysis are derived and summarized in this appendix. We present the general forms of the momentum integrals appearing in the calculation, including those involving arbitrary powers of $p$, together with additional auxiliary integrals required for evaluating the number and current densities. These derivations explicitly demonstrate how modified Bessel functions arise naturally in the final expressions, thereby providing a systematic foundation for the analytic results used throughout the main text. 
The resulting expressions are consistent with standard integral identities reported in refs.~\cite{gradshteyn2007table,abramowitz1964handbook}. We have
\begin{align}
    \int_{0}^{\infty}\frac{d^{3}p}{(2\pi)^{3}}p^{n}f_{eq}=\frac{2^{1+\frac{n}{2}}}{2\pi^{2}}\frac{\Gamma\left(\frac{3+n}{2}\right)}{\sqrt{\pi}}m^{3+n}\sum_{j=1}^{\infty}(-1)^{j+1}e^{j\frac{\mu}{T}}\left(\frac{jm}{T}\right)^{-1-\frac{n}{2}}K_{\frac{4+n}{2}}\left(\frac{jm}{T}\right),
\end{align}
\begin{align}
    \int_{0}^{\infty}\frac{d^{3}p}{(2\pi)^{3}}p^{n}D_{eq}=\frac{2^{1+\frac{n}{2}}}{2\pi^{2}}\frac{\Gamma(\frac{3+n}{2})}{\sqrt{\pi}}m^{3+n}\sum_{j=1}^{\infty}(-1)^{j+1}je^{j\frac{\mu}{T}}\left(\frac{jm}{T}\right)^{-1-\frac{n}{2}}K_{\frac{4+n}{2}}\left(\frac{jm}{T}\right),
\end{align}
\begin{align}
    \int_{0}^{\infty}\frac{d^{3}p}{(2\pi)^{3}}p^{n}D_{eq}(1-f_{eq,f})=\frac{2^{1+\frac{n}{2}}}{2\pi^{2}}\frac{\Gamma(\frac{3+n}{2})}{\sqrt{\pi}}m^{3+n}\sum_{j=1}^{\infty}(-1)^{j+1}j^{2}e^{j\frac{\mu}{T}}\left(\frac{jm}{T}\right)^{-1-\frac{n}{2}}K_{\frac{4+n}{2}}\left(\frac{jm}{T}\right),
\end{align}
\begin{align}
\begin{split}
    &\int_{0}^{\infty}\frac{d^{3}p}{(2\pi)^{3}}p^{n}D_{eq}(1-6f_{eq,f}+6f_{eq,f}^{2})\\
   &\qquad=\frac{2^{1+\frac{n}{2}}}{2\pi^{2}}\frac{\Gamma(\frac{3+n}{2})}{\sqrt{\pi}}m^{3+n}\sum_{j=1}^{\infty}(-1)^{j+1}j^{3}e^{j\frac{\mu}{T}}\left(\frac{jm}{T}\right)^{-1-\frac{n}{2}}K_{\frac{4+n}{2}}\left(\frac{jm}{T}\right),\\
    \end{split}
\end{align}
\begin{align}
&\int_{b}^{\infty} x^{r} (x^{2} - b^{2})^{n} e^{-a x} \, dx\\
&= \frac{1}{2} b^{2n} \Bigg[
\frac{
b^{1+m} \, \Gamma\left(-\frac{1}{2} - \frac{r}{2} - n\right) \, \Gamma(1 + n) \,
{}_1F_2\left(\frac{1}{2} + \frac{r}{2}; \frac{1}{2}, \frac{3}{2} + \frac{r}{2} + n; \frac{a^2 b^2}{4}\right)
}
{
\Gamma\left(\frac{1}{2} - \frac{r}{2}\right)
} \notag \\
&\qquad\qquad -
\frac{
a \, b^{2 + r} \, \Gamma\left(-1 - \frac{r}{2} - n\right) \, \Gamma(1 + n) \,
{}_1F_2\left(1 + \frac{r}{2}; \frac{3}{2}, 2 + \frac{r}{2} + n; \frac{a^2 b^2}{4}\right)
}
{
\Gamma\left(-\frac{r}{2}\right)
} \notag \\
&\qquad\qquad +
2 a^{-1 - r} (ab)^{-2n} \, \Gamma(1 + r + 2n) \,
{}_1F_2\left(-n; -\frac{r}{2} - n, \frac{1}{2} - \frac{r}{2} - n; \frac{a^2 b^2}{4}\right)
\Bigg]~\mathrm{if}~r>-1,
\end{align}
${}_1F_2$ is a generalized hypergeometric function.
\begin{align}
\begin{split}
\int_b^\infty {x \left(x^2 - b^2\right)^{n - \frac{3}{2}}}{e^{-a x}} \, dx =\frac{2^{n - 1} \, b^n \, \Gamma\left(n - \frac{1}{2} \right)}{\sqrt{\pi}} K_n(a b),
\quad \text{for } n > \frac{1}{2},\ b > 0,\ a > 0
\end{split}
\end{align}

\begin{align}
    \int_{b}^{\infty} x(x^{2}-b^{2})^{n-1} e^{-ax}dx=\frac{2^{-\frac{1}{2} + n} \left( \frac{b}{a} \right)^n \sqrt{a b} \, K_{\frac{1}{2} + n}(a b) \, \Gamma(n)}{\sqrt{\pi}},
\end{align}
\begin{align}
    \int_{b}^{\infty} (x^{2}-b^{2})^{n-1} e^{-ax}dx=\frac{2^{-\frac{1}{2} + n} \left( \frac{b}{a} \right)^{-\frac{1}{2} + n} \, K_{-\frac{1}{2} + n}(a b) \, \Gamma(n)}{\sqrt{\pi}}.
\end{align}
The modified Bessel function of the second kind admits the integral representation
\begin{equation}
K_n\left(\frac{jm}{T}\right)=\int_0^{\infty}\cosh(n\theta)e^{-\frac{jm}{T}\cosh\theta}d\theta.
\end{equation}
Similarly, we define the generalized integral
\begin{equation}
K_{\mathcal{I},n}\left(\frac{jm}{T}\right)=\int_0^{\infty}e^{-\frac{jm}{T}\cosh\theta}\frac{d\theta}{(\cosh\theta)^n}.
\end{equation}

\section{Reduction to the Massless Limit}\label{Mass-to-Massless Conversion}
In this appendix, we examine the massless limit of the obtained expressions. As seen from Eqs.~\eqref{J Total} and \eqref{n_total}, the particle mass appears both explicitly and implicitly through the modified Bessel functions. To study the limit $m \to 0$, we expand the Bessel functions in their small-argument form and analyze the resulting expressions. Below we outline the procedure used to obtain the massless limits of the relevant quantities:

\begin{align}
    K_1\left(\frac{jm}{T}\right)=\frac{T}{j m}+\frac{j m \left(2 \log \left(\frac{j m}{2 T}\right)+2 \gamma -1\right)}{4 T}+\frac{j^3 m^3 \left(4 \log \left(\frac{j m}{2 T}\right)+4 \gamma -5\right)}{64 T^3}+O\left(m^5\right),
\end{align}
\begin{align}
\begin{split}
   K_2\left(\frac{jm}{T}\right)=&\frac{2 T^2}{j^2 m^2}-\frac{1}{2}-\frac{m^2 \left(j^2 \left(4 \log \left(\frac{j m}{2 T}\right)+4 \gamma -3\right)\right)}{32 T^2}\\
   &~~~~~~~~~~~-\frac{m^4 \left(j^4 \left(12 \log \left(\frac{j m}{2 T}\right)+12 \gamma -17\right)\right)}{1152 T^4}+O\left(m^5\right),
   \end{split}
\end{align}
\begin{align}
   K_3\left(\frac{jm}{T}\right)= \frac{8 T^3}{j^3 m^3}-\frac{T}{j m}+\frac{j m}{8 T}+\frac{j^3 m^3 \left(12 \log \left(\frac{j m}{2 T}\right)+12 \gamma -11\right)}{576 T^3}+O\left(m^5\right).
\end{align}
The above expressions provide the small-argument expansions of the modified Bessel functions $K_{3}(z)$, $K_{2}(z)$ and $K_{1}(z)$ for $z = jm/T \ll 1$. Here $\gamma$ is Euler-Mascheroni constant. These expansions are important because the Bessel functions appearing in the number and current densities become increasingly divergent in the limit $m \to 0$, with the leading behaviors $K_{3}(z) \sim z^{-3}$, $K_{2}(z) \sim z^{-2}$, and $K_{1}(z) \sim z^{-1}$.
Writing them in series form makes these divergences explicit and isolates the mass-dependent contributions. 
This allows one to identify how the explicit powers of $m$ in the integrals regulate the apparent singularities and yield a well-defined massless limit. We have
\begin{align}\label{d1}
    mK_1\left(\frac{jm}{T}\right)=\frac{T}{j}+\frac{j m^2 \left(2 \log \left(\frac{j m}{2 T}\right)+2 \gamma -1\right)}{4 T}+\frac{j^3 m^4 \left(4 \log \left(\frac{j m}{2 T}\right)+4 \gamma -5\right)}{64 T^3}+O\left(m^5\right),
\end{align}
\begin{align}\label{d2}
    m^2 K_2 \left(\frac{jm}{T}\right)=\frac{2T^2}{j^2}-\frac{m^2}{2}-\frac{m^4 \left(j^2 \left(4 \log \left(\frac{j m}{2 T}\right)+4 \gamma -3\right)\right)}{32 T^2}-O(m^5),
\end{align}
\begin{align}\label{d3}
    m^3 K_3\left(\frac{jm}{T}\right)=\frac{8 T^3}{j^3}-\frac{m^2 T}{j}+\frac{j m^4}{8 T}+O\left(m^5\right).
\end{align}
These Eqs. \eqref{d1}, \eqref{d2} and \eqref{d3}, illustrate how the explicit powers of the mass multiplying the Bessel functions combine with their small-argument expansions to cancel the divergent terms. 
In particular, multiplying $K_{2}(jm/T)$ by $m^{2}$ removes the leading $m^{-2}$ divergence, while multiplying $K_{1}(jm/T)$ by $m$ eliminates the $m^{-1}$ pole. Likewise, multiplying $K_{3}(jm/T)$ by $m^{3}$ cancels the $m^{-3}$ divergence, leaving finite contributions in the limit $m \to 0$. 
These cancellations reflect the structure of the relativistic phase-space integrals and ensure that the massless limit of the number and current densities is obtained by retaining only the finite, mass-independent terms. We obtain,
\begin{align}\label{jsink1}
\frac{mT}{\pi^2}\sum_{j=1}^{\infty}\frac{(-1)^{j+1}}{j}\sinh{\left(\frac{j\mu}{T}\right)} K_1 \left(\frac{jm}{T}\right) \xrightarrow[\;m=0]{} \frac{T^2}{2\pi^2}\left[\text{Li}_2\left\{-\exp\left(-\frac{\mu}{T}\right)\right\}-\text{Li}_2\left\{-\exp\left(\frac{\mu}{T}\right)\right\}\right],
\end{align}
\begin{align}\label{jcosk1}
\frac{mT}{\pi^2}\sum_{j=1}^{\infty}\frac{(-1)^{j+1}}{j}\cosh{\left(\frac{j\mu}{T}\right)} K_1 \left(\frac{jm}{T}\right) \xrightarrow[\;m=0]{} -\frac{T^2}{2\pi^2}\left[\text{Li}_2\left\{-\exp\left(-\frac{\mu}{T}\right)\right\}+\text{Li}_2\left\{-\exp\left(\frac{\mu}{T}\right)\right\}\right],
\end{align}
\begin{align}\label{jsin}
\frac{m^{2}T}{\pi^2}\sum_{j=1}^{\infty}\frac{(-1)^{j+1}}{j}\sinh{\left(\frac{j\mu}{T}\right)} K_2 \left(\frac{jm}{T}\right) \xrightarrow[\;m=0]{} \frac{T^3}{\pi^2}\left[\text{Li}_3\left\{-\exp\left(-\frac{\mu}{T}\right)\right\}-\text{Li}_3\left\{-\exp\left(\frac{\mu}{T}\right)\right\}\right],
\end{align}
\begin{align}\label{jcos}
\frac{m^{2}T}{\pi^2}\sum_{j=1}^{\infty}\frac{(-1)^{j+1}}{j}\cosh{\left(\frac{j\mu}{T}\right)} K_2 \left(\frac{jm}{T}\right) \xrightarrow[\;m=0]{} -\frac{T^3}{\pi^2}\left[\text{Li}_3\left\{-\exp\left(-\frac{\mu}{T}\right)\right\}+\text{Li}_3\left\{-\exp\left(\frac{\mu}{T}\right)\right\}\right],
\end{align}
\begin{align}\label{0sin}
\frac{m^{2}T}{\pi^2}\sum_{j=1}^{\infty}(-1)^{j+1}\sinh{\left(\frac{j\mu}{T}\right)} K_2 \left(\frac{jm}{T}\right) \xrightarrow[\;m=0]{} \frac{T^3}{\pi^2}\left[\text{Li}_2\left\{-\exp\left(-\frac{\mu}{T}\right)\right\}-\text{Li}_2\left\{-\exp\left(\frac{\mu}{T}\right)\right\}\right],
\end{align}
\begin{align}\label{0cos}
\frac{m^{2}T}{\pi^2}\sum_{j=1}^{\infty}(-1)^{j+1}\cosh{\left(\frac{j\mu}{T}\right)} K_2 \left(\frac{jm}{T}\right) \xrightarrow[\;m=0]{} -\frac{T^3}{\pi^2}\left[\text{Li}_2\left\{-\exp\left(-\frac{\mu}{T}\right)\right\}+\text{Li}_2\left\{-\exp\left(\frac{\mu}{T}\right)\right\}\right],
\end{align}

\begin{align}\label{j^2 cos}
\frac{m^2T}{\pi^2}\sum_{j=1}^{\infty}\frac{(-1)^{j+1}}{j^2}\cosh{\left(\frac{j\mu}{T}\right)} K_2 \left(\frac{jm}{T}\right) \xrightarrow[\;m=0]{} -\frac{T^3}{\pi^2}\left[\text{Li}_4\left\{-\exp\left(-\frac{\mu}{T}\right)\right\}+\text{Li}_4\left\{-\exp\left(\frac{\mu}{T}\right)\right\}\right],
\end{align}
\begin{align}\label{K3 cos}
\frac{m^{3}T}{\pi^2}\sum_{j=1}^{\infty}\frac{(-1)^{j+1}}{j}\cosh{\left(\frac{j\mu}{T}\right)} K_3 \left(\frac{jm}{T}\right) \xrightarrow[\;m=0]{} -\frac{4T^4}{\pi^2}\left[\text{Li}_4\left\{-\exp\left(-\frac{\mu}{T}\right)\right\}+\text{Li}_4\left\{-\exp\left(\frac{\mu}{T}\right)\right\}\right],
\end{align}
\begin{equation}
\frac{m^{3} T}{\pi^{2}} \sum_{j=1}^{\infty} (-1)^{j+1} \sinh\left(\frac{j\mu}{T}\right) K_{3}\left(\frac{jm}{T}\right)\xrightarrow[\;m=0]{}\frac{4T^4}{\pi^2}\left[\text{Li}_3\left\{-\exp\left(-\frac{\mu}{T}\right)\right\}-\text{Li}_3\left\{-\exp\left(\frac{\mu}{T}\right)\right\}\right].
\end{equation}

\bibliographystyle{apsrev4-1}
\bibliography{ref.bib}

\end{document}